\documentclass[modern]{aastex62}

\received{May 10, 2018}
\accepted{July 10, 2018}
\submitjournal{AJ}

\shorttitle{Main Belt Asteroid Shape Distribution from Gaia DR2}
\shortauthors{Mommert et al.}
\begin{document}

\title{The Main Belt Asteroid Shape Distribution from Gaia Data Release 2}

\correspondingauthor{Michael Mommert}
\email{michael.mommert@lowell.edu}

\author[0000-0002-8132-778X]{Michael Mommert}
\affil{Lowell Observatory, 1400 W.\ Mars Hill Rd., Flagstaff, AZ, 86001, USA}

\author{Andrew McNeill}
\affiliation{Northern Arizona University, Department of Physics and Astronomy, Flagstaff, AZ 86011, USA}

\author{David E. Trilling}
\affiliation{Northern Arizona University, 
Department of Physics and Astronomy,
Flagstaff, AZ 86011, USA}

\author{Nicholas Moskovitz}
\affil{Lowell Observatory, 1400 W.\ Mars Hill Rd., Flagstaff, AZ, 86001, USA}

\author{Marco Delbo'}
\affiliation{Universit\'{e} C\^{o}te d'Azur, 
CNRS Lagrange, Observatoire de la C\^{o}te d'Azur, 
CS 34229 - F 06304 NICE Cedex 4, France}



\begin{abstract}
  Gaia Data Release 2 includes observational data for
  14,099~pre-selected asteroids. From the sparsely sampled $G$ band
  photometry, we derive lower-limit lightcurve amplitudes for
  11,665 main belt asteroids in order to provide constraints on the
  distribution of shapes in the asteroid main belt. Assuming a
  triaxial shape model for each asteroid, defined through the axial
  aspect ratios $a > b$ and $b=c$, we find an average $b/a=0.80\pm0.04$
  for the ensemble, which is in agreement with previous results. By
  combining the Gaia data with asteroid properties from the
  literature, we investigate possible correlations of the aspect ratio
  with size, semi-major axis, geometric albedo, and intrinsic
  color. Based on our model simulations, we find that main belt
  asteroids greater than 50~km in diameter on average have higher
  $b/a$ aspect ratios (are rounder) than smaller asteroids.  We
  furthermore find significant differences in the shape distribution
  of main belt asteroids as a function of the other properties that do
  not affect the average aspect ratios. We conclude that a more
  detailed investigation of shape distribution correlations requires a
  larger data sample than is provided in Gaia Data Release 2.
\end{abstract}



\section{Introduction} 
\label{sec:intro}

Asteroid physical properties provide important constraints on the
formation and evolution of our Solar System, including clues to the
different phases of orbital instability and migration of the giant
planets \citep[see, e.g.,][]{Morbidelli2015}.  Collisions within the
asteroid belt can play a major role altering the asteroid population.
These clues are recorded in the shapes of individual asteroids, so
that a global understanding of the shape distribution --- as informed
by a large catalog --- provides insight into our Solar System's
evolution.

Shape information can be determined for individual asteroids from
photometric data over a long period of time, with Doppler-delay radar
imaging, with ground-based adaptive optics techniques, or {\it in
  situ} via spacecraft data. These last three techniques are limited
in the number of targets that can be reasonably observed. As a result,
only a small percentage of km-scale main belt asteroids have known
shapes and spin-pole orientations. For instance, the DAMIT
database\footnote{Database of Asteroid Models from Inversion
  Techniques:
  \url{http://astro.troja.mff.cuni.cz/projects/asteroids3D/web.php}}
\citep{Durech2010} holds information for less than 1,000 asteroids as
of writing this. Currently, the predominant method for determination
of asteroid shape and spin properties is the inversion of dense
photometric lightcurves, as initially developed by
\citet{Kaasalainen2001a} and \citet{Kaasalainen2001b}. Without dense
photometric data it is difficult to derive detailed shape and spin
pole orientations for individual asteroids. With a sufficiently large
dataset, however, it is possible to determine a statistical shape
distribution, even if only a small number of data points are present
for each individual asteroid. The advantage of a large untargeted
dataset is that it provides an estimate for a population's shape
distribution without being subject to observational biases in favor of
elongated objects (i.e.,~higher lightcurve amplitudes) that are
commonly present in targeted observations. Previous work on the shape
distribution for main belt asteroids (MBAs) has been carried out by
\citet{McNeill2016}, \citet{Cibulkova2016}, \citet{Nortunen2017}, and
\citet{Cibulkova2018}.

European Space Agency's Gaia astrometric space observatory
\citep{Gaia2016} measures the positions, distances, proper motions,
and other physical properties of more than one billion stars in our
galaxy with unprecedented accuracy. In addition to its main mission,
Gaia will also observe a significant fraction of the currently known
asteroid population (${\sim}$350,000 Solar System small bodies) and
discover previously unknown asteroids \citep{Spoto2018}.  The sample
of Gaia-observed asteroids provides a large uniform and
magnitude-limited set of observations perfectly suited to
independently investigate the distribution of asteroid shapes. 
% By
% cross-matching the sample with other resources, we can investigate
% trends and correlations with other physical parameters, including
% intrinsic color, diameter, and surface albedo, as well as orbital
% properties, all of which provide constraints on the long-term
% evolution of our Solar System.

% Here we present the aggregate shape distribution of 11,689 main belt
% asteroids from Gaia Data Release 2 (DR2). In
% Section~\ref{sec:analysis} we describe our data analysis procedure
% that results in estimated lightcurve amplitudes for individual
% Gaia-observed asteroids. In Section~\ref{sec:results} we present our
% resulting overall derived shape distribution for the entire Gaia DR2
% moving object catalog as well as comparisons of this shape
% distribution with colors, albedos, diameters, and semi-major axis.  In
% Section~\ref{sec:discussion} we demonstrate our analysis validation
% and discuss our results in the context of previous work.

\section{Gaia Data Release 2}
\label{sec:gaia}

%As a result of Gaia's observing mode that
%repeatedly scans the entire sky, the spacecraft also detects
%point-like Solar System objects, the vast majority of which are
%asteroids. 
Gaia Data Release~2 \citep[DR2,][]{Spoto2018} includes 1,977,702
astrometric observations 
% (provided in data table {\tt ssoobs}, naming
% conventions used here according to CDS
% Vizier\footnote{\url{http://vizier.u-strasbg.fr/viz-bin/VizieR-3?-source=I/345/ssoobs}})
of 14,099 pre-selected asteroids, observed between 5 August 2014 and
23 May 2016.  Solar System moving objects are identified through
association with known asteroids with well-defined orbits. 
% Gaia is
% designed to detect point-like sources, severely limiting its ability
% to detect extended sources (e.g., active comets) and trailed,
% fast-moving asteroids. 
The majority of DR2 asteroids are main belt asteroids, but the sample
also includes a small number of Near-Earth Asteroids, Jupiter
Trojans, and trans-Neptunian objects. Each ``transit'' of a target
across the detector array leads to 9 individual detections across a
typical time span on the order of 40~s. Target positions and epochs
% in {\tt ssoobs} 
are provided for individual detections, whereas
photometric information is averaged per transit.
%Each Solar System
%object detection in DR2 is associated with an already known Solar
%System object and the official IAU number identifier as provided by the Minor Planet Center are provided in {\tt ssoobs}; all objects in the catalog have been unanimously identified.
Photometric information on asteroids in DR2 is limited to Gaia $G$ band (0.33--1.0~$\mu$m)
magnitudes, fluxes, and flux uncertainties.

\section{Data Analysis}
\label{sec:analysis}

\subsection{Data Preparation}
\label{sec:preparation}

All DR2 asteroid observations were downloaded from CDS Vizier\footnote{\url{http://vizier.u-strasbg.fr/viz-bin/VizieR}}. Since the photometry
per object is the average of all individual observations within a
single transit, we extract only the midtime epoch from each transit.
The maximum number of transits for any individual object is 53 and the
median number of transits per target is 9. We require a minimum of 5
transits per object to be considered in this analysis. While this
  criterion is somewhat arbitrary, it allows for a meaningful
  statistical analysis of the data set, which is not guaranteed
  for a smaller number of observations. For each transit, we derive
approximate $G$ band magnitude uncertainties from the provided fluxes
and flux uncertainties.
%Photometric information is not available for 118 asteroids in DR2. 
We obtain ephemerides for the respective epochs and relative to Gaia
using the {\tt astroquery.jplhorizons}
module,
which queries ephemerides from the JPL Horizons system
\citep{Giorgini1996}. For each $G$ band observation, we derive the
reduced $G$ band magnitude, $G_{11}(\alpha)$, which corrects for
geometric effects as a result of the heliocentric distance of the
target, $r$, and the distance from the observer, $\Delta$, but
preserves the solar phase angle ($\alpha$) dependence of the
brightness: $G_{11}(\alpha) = G - 5 \log_{10} r \Delta$. We remove all
objects that are not considered main belt asteroids from our sample
since the population sizes of near-Earth asteroids, Jupiter Trojans,
and trans-Neptunian objects in DR2 are insufficient for this
analysis. Our final sample consists of 121,819 $G$ band photometric
observations of 12,059 different main belt asteroids.

\subsection{Phase Correction and Amplitude Measurement}
\label{sec:phasecurve}

Variations in the $G_{11}(\alpha)$ magnitudes are the result of
lightcurve variations due to the irregular shapes of asteroids and
their rotation, as well as solar phase angle effects \cite[see,
e.g.,][]{Bowell1989}. In order to extract shape information from the
data, the phase angle effect has to be corrected for. On average, DR2
observations span a phase angle range of 6\degr\ around an average
phase angle of 19\degr, which is insufficient to perform a reliable
phase curve fit. In order to correct for solar phase angle effects on
the Gaia photometry, we instead use existing phase curve information
on our targets from 
% the Minor Planet
% Center\footnote{\url{https://minorplanetcenter.net/iau/MPCORB.html}}
\citet{Veres2015} and \citet{Oszkiewicz2011}. While \citet{Veres2015}
only used photometric observations obtained by Pan-STARRS1
\citep{Kaiser2010}, \citet{Oszkiewicz2011} compiled low-precision
photometric data from the Minor Planet
Center\footnote{\url{https://minorplanetcenter.net/}} (MPC). Both
sources make use of the $H$-$G_{12}$ two-parameter phase function
\citep{Muinonen2010} and provide measured absolute magnitudes ($H$)
and slope parameters ($G_{12}$) with corresponding
uncertainties. Preference is given to slope parameters reported by
\citet{Veres2015} if their measurement is based on at least 5
observations; this selection is based on the improved reliability
  and existence of photometric uncertainties in the Pan-STARRS1 data
  in contrast to the MPC data. If no measurement from
\citet{Veres2015} is available, we adopt the slope parameter reported
by \citet{Oszkiewicz2011}. 
% Despite known issues with the H-G phase curve formalism
% \citep[e.g.,][]{Muinonen2010} we make use of it as it represents the
% largest combined catalog of asteroid phase curve information.  ,
% since these measurements are based on more precise observations than
% observations reported to the Minor Planet Center.
In our final sample, 56\% of the $G_{12}$ slope parameters were taken
from \citet{Veres2015} and 44\% were taken from
\citet{Oszkiewicz2011}. Only 24 targets in our sample had no
measurement of the slope parameter $G_{12}$ from neither
\citet{Veres2015} nor \citet{Oszkiewicz2011}; these targets were
rejected from our further analysis.

\begin{figure}
\epsscale{0.8}
\plotone{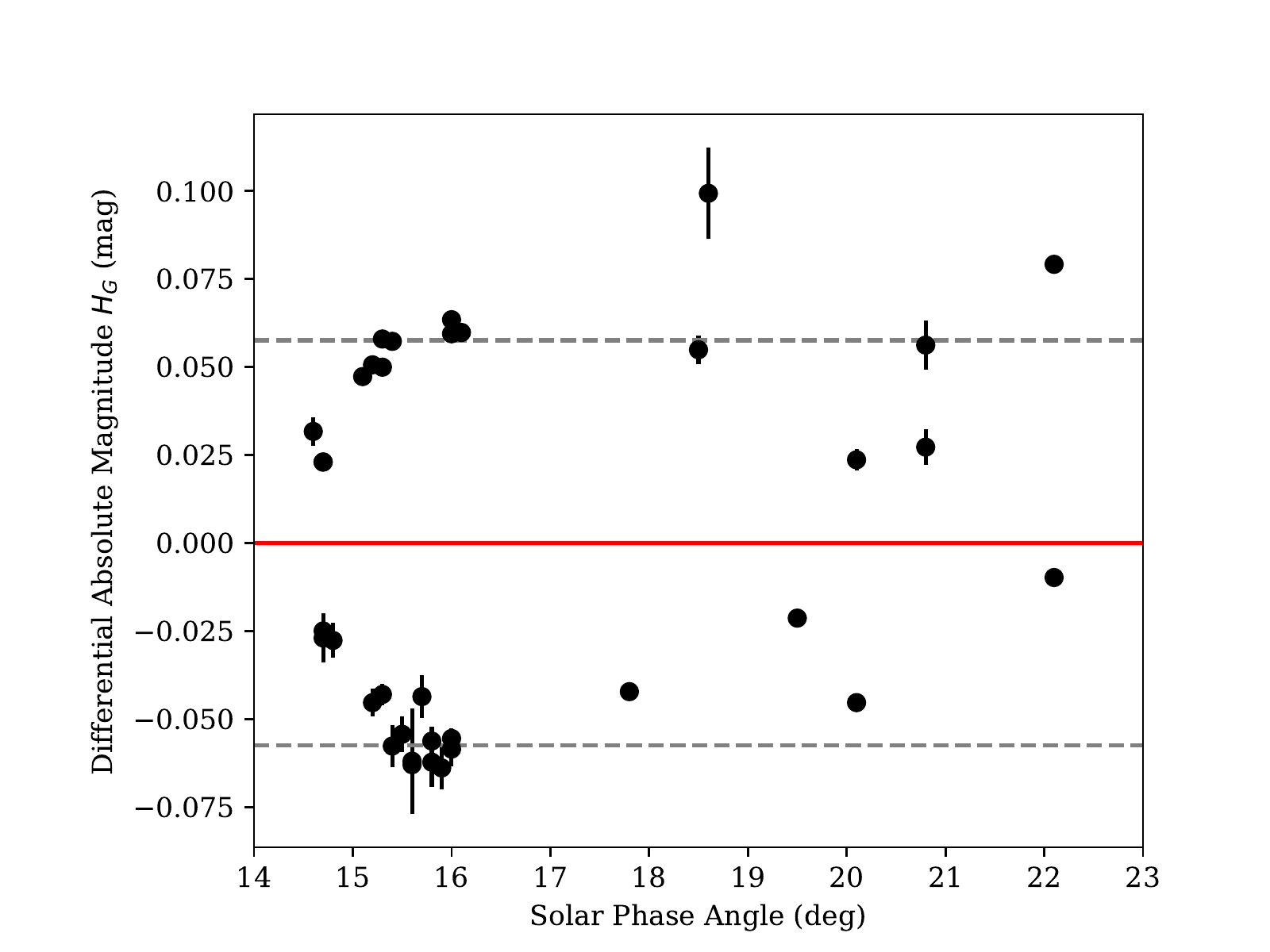}
\caption{Example lightcurve amplitude derivation for asteroid (79)
  Eurynome. The diagram shows differential absolute magnitudes in $G$
  band ($H_G$) derived using the method defined in Section
  \ref{sec:phasecurve} and plotted against solar phase angle. These
  differential magnitudes are obtained by subtracting the weighted
  average $H_G$ from the individual measurements to better visualize
  the scatter around the target's normalized average absolute
  magnitude (red line). We drive the lower limit lightcurve amplitude
  (peak-to-peak) as two times the standard deviation (dashed lines) of
  this scatter. In the case of Eurynome, we derive a lower limit
  amplitude of 0.10~mag from 36 DR2 observations. The Asteroid
  Lightcurve Database \citep{Warner2009} lists a maximum lightcurve
  amplitude of 0.25~mag, which is consistent with our lower limit. The
  sloping of the $H_G$ magnitudes as a function of solar phase angles
  suggests that the adopted $G_{12}$ photometric slope parameter does
  not describe the target's phase curve well in this particular
  case. The linear regression analysis for this data set yields a 
    coefficient of determination parameter $r^2 = 0.003$ and a
    $p$-value of 0.75, which does not lead to a rejection of this object
    from our sample (see Section
  \ref{sec:phasecurve}).  \label{fig:analysis_example}}
\end{figure}

Based on the solar phase angles during the Gaia observations 
%as provided from JPL Horizons, 
and the slope parameter $G_{12}$ from the literature, we correct the
$G_{11}(\alpha)$ magnitudes for phase angle effects, resulting in
normalized $H_G$ absolute magnitudes that reveal the targets'
lightcurve variations. In order to account for uncertainties in the
derivation of the slope parameter $G_{12}$, we adopt a Monte Carlo
approach in which we vary $G_{12}$ using a Gaussian model based on the
$G_{12}$ uncertainties reported by  \citet{Veres2015} or
  \citet{Oszkiewicz2011}; the latter source utilizes asymmetric
uncertainties, which we adopt throughout this work.
%  distribution derived by \cite{Veres2015}. For each target
% we choose a random uncertainty drawn from a Gaussian distribution
% centered on 0.3 with a standard deviation of 0.1 \citep[compare to
% Figure 9 by][]{Veres2015}. Based on this randomized uncertainty, 
We
vary $G_{12}$ using two different Gaussian distributions corresponding to the upper and lower 1$\sigma$ uncertainties, re-derive the magnitude offsets and $H_G$
values, and calculate the weighted standard deviation (we use the
inverse square uncertainties of the Gaia photometry as weights) of the
resulting $H_G$ magnitudes. We adopt the average of the weighted $H_G$
standard deviations from 1,000 Monte Carlo runs as the target's
lightcurve amplitude. This method tends to underestimate large lightcurve
amplitude, especially in the case of sparse data; our derived
amplitudes $A$ hence represent lower limits of the real lightcurve
amplitudes (see Section \ref{sec:validation} for a discussion). Figure
\ref{fig:analysis_example} shows an example of the method that we use
to derive lightcurve amplitudes. Finally, 
% we reject 
% lightcurve
% amplitudes greater than 1~mag in a visual inspection as those might be
% subject to inaccurate slope parameters corrupting the results of this
% method.
we use a linear regression model ({\tt scipy.stats.linregress})
  to identify objects with strong linear slopes in their $H_G$
  magnitudes as a function of solar phase angle, which are most likely
  caused by incorrect phase slope parameters $G_{12}$. Objects with
  coefficient of determination parameters $r^2 > 0.5$ or $p$-values
  less than 0.05 are rejected as they unanomously point to severe
  sloping. These criteria are somewhat arbitrary and result from the
  visual inspection of a few hundred objects. However, they provide a
  conservative rejection scheme.

\section{Results}
\label{sec:results}

For each DR2 target, we derive the lower limit lightcurve amplitude
using the method introduced in Section \ref{sec:analysis}. The results
for all 11,665 main belt asteroids for which this method
succeeded (84.3\% of the complete DR2 sample; 98.6\% of the main
  belt asteroids in the sample with sufficient data as defined in
  Section \ref{sec:preparation}) are listed in
Table~\ref{tab:results}.

% num   amp  n alpl alpu    G12  Guu   Gul  s     a*      a   diam    pv

\begin{deluxetable*}{lcccccccccccc}
\tablecaption{DR2 Main Belt Asteroid Lightcurve Amplitudes and Auxiliary Data \label{tab:results}}
\tablecolumns{9}
\tablewidth{0pt}
\tablehead{
\colhead{MPC} &
\colhead{Amp.\tablenotemark{a}} &
\colhead{$N_{\mathrm{obs}}$\tablenotemark{b}} &
\multicolumn2c{Phase Angle\tablenotemark{c}} &
\multicolumn4c{Slope Parameter\tablenotemark{d}} & 
\colhead{SDSS Color\tablenotemark{e}} &
\colhead{$a$\tablenotemark{f}} &
\colhead{Diameter\tablenotemark{g}} &
\colhead{Albedo\tablenotemark{h}}\\[-5pt]
\colhead{Number} & 
\colhead{(mag)} &
\colhead{} & 
\colhead{Min (\degr)} & 
\colhead{Max (\degr)} &
\colhead{$G_{12}$} &
\colhead{$\sigma_+$} &
\colhead{$\sigma_-$} &
\colhead{Source} & 
\colhead{$a^{\text{*}}$} &
\colhead{(au)} &
\colhead{(km)} &
\colhead{$p_V$}
}
\startdata
8 & 0.15 & 8 & 17.2 & 28.2 & 0.33 & 0.08 & -0.08 & o & ... & 2.201 & 158.4 & 0.21 \\
33 & 0.11 & 26 & 12.5 & 26.3 & 0.15 & 0.53 & -0.53 & v & ... & 2.876 & 50.9 & 0.24 \\
280 & 0.11 & 27 & 14.3 & 22.3 & 0.38 & 0.31 & -0.31  & v & -0.04 & 2.942 & 55.8 & 0.03 \\
460 & 0.10 &  8 & 17.2 & 20.7 & 0.32 & 0.11 & -0.11 & o & 0.09 & 2.717 & 19.7 & 0.26 \\
11644 & 0.13 &  7 & 12.2 & 17.4 & -0.41 & 0.31 & -0.31  & v  & 0.01 & 3.118 & 13.4 & 0.09 \\
... & ... & ... & ... & ... & ... & ... & ...& ... & ... & ...\\
\enddata
\tablenotetext{a}{Lightcurve amplitude lower limit as derived with the method defined in Section \ref{sec:analysis}. We provide the nominal peak-to-peak value.}

\tablenotetext{b}{Number of observations in DR2 with photometric measurements used in this analysis; all objects with $N_{\mathrm{obs}} < 5$ have been removed from this sample.}

\tablenotetext{c}{Solar phase angle range over which this object has been observed in degrees. We provide the minimum ({\em Min}) and maximum ({\em Max}) solar phase angle as provided in DR2.}

\tablenotetext{d}{Adopted photometric slope parameter $G_{12}$ as defined by \citep{Muinonen2010}. We provide the adopted value ($G_{12}$) with corresponding upper ($\sigma_+$) and lower ($\sigma_-$) $1\sigma$ uncertainties, and the source of these values ({\em Source}): ``{\em o}'' refers to \citet{Oszkiewicz2011} and ``{\em v}'' refers to \citet{Veres2015}.}

\tablenotetext{e}{Color parameter $a^{\text{*}}$ as defined by \citet{Ivezic2001} and obtained from the Sloan Digitized Sky Survey Moving Object Catalog \citep{Ivezic2005}.}

\tablenotemark{f}{Semi-major axis of the target as obtained from the Minor Planet Center database in astronomical units.}

\tablenotemark{g}{Asteroid diameter as derived by \citet{Mainzer2016} in kilometers.}

\tablenotemark{h}{Asteroid geometric albedo ($V$ band) as derived by \citet{Mainzer2016}.}

\tablecomments{This table is only a representative excerpt and is published 
in its entirety in the machine readable format.  A portion is
shown here for guidance regarding its form and content.}
\end{deluxetable*}

\subsection{Asteroid Shape Distribution}
\label{sec:results_shapes}

In order to derive the average shape distribution of main belt
asteroids, we implement the model used by \citet{McNeill2016}. This
model generates a synthetic population of asteroids with random shapes
and spin pole orientations and compares them to the observed data
using a $\chi^2$ goodness of fit test. Shapes are generated assuming a
triaxial ellipsoidal shape with axis ratios $b/a$ and $b/c$; we
generate random ratios $0.2 \leq b/a < 1.0$ and presume $b=c$. We
assume the spin pole latitudes and longitudes of the observed objects
based on the distribution for small (diameter ${\leq}30$~km)
  asteroids obtained by \citet{Hanus2011}; spin frequencies are
derived from a uniform distribution covering 1--10.9~day$^{-1}$,
corresponding to rotational periods ranging from the spin barrier at
2.2~h to a period of 24~hr. This assumption is consistent with the
flat distribution of measured rotational frequencies at small sizes
\citep{Pravec2002}.  We account for phase angle variations in each
target by assigning monotonically increasing or decreasing phase angle
values based on the ensemble properties of DR2 targets (see Section
\ref{sec:phasecurve} and Table \ref{tab:results}). We furthermore
account for the fact that lightcurve amplitudes measured as part of
this work represent lower limits by increasing each measured amplitude
by 51\% and adding a noise contribution following a normal
distribution with a relative standard deviation of 20\% in each Monte
Carlo run. This approach is justified in Section \ref{sec:validation}.

Based on 10,000 Monte Carlo runs using 11,665 synthetic objects each,
we find an average aspect ratio for main belt asteroids
$b/a=0.80\pm 0.04$. The nominal value is defined as the
global minimum $\chi^2$ of all model runs; the uncertainty is derived
within a reasonable range ($\chi^2 \leq 2$) around the global minimum of the
distribution. We discuss this result in Sections \ref{sec:previouswork} and \ref{sec:limitations}.

\subsection{Trends and Correlations}
\label{sec:results_trends}

We investigate potential correlations between our derived lightcurve
amplitudes and intrinsic target properties: semi-major axis, color,
albedo, and diameter. 
% We emphasize that amplitude solutions for
% individual asteroids may not be reliable but that globally our
% technique correctly assesses the amplitude distribution, as described
% in Section~\ref{sec:validation}.  
For this analysis, we match our sample target list with 
% the objects listed in Table \ref{tab:results} with 
previously derived asteroid color parameters $a^{\text{*}}$
\citep{Ivezic2001} from the Sloan Digital Sky Survey (SDSS) Moving
Object Catalog \citep[Data Release 3,][]{Ivezic2005}, diameters and
albedos of WISE-observed small bodies \citep{Mainzer2016}, as well as
semi-major axes as provided by the Minor Planet Center database. 
  We use {\tt sbpy.data.Names.parse\_asteroid} to parse the respective
  asteroid identifiers in the different archives to perform the object
  matching. The SDSS color parameter $a^{\text{*}}$ is derived as the
average over all available measurements; in order to reject outliers,
we restrict ourselves to the range $-0.3 < a^{\text{*}} <
0.3$.
Similarly, WISE diameters and albedos are derived as averages if
multiple observations are available.

\begin{figure}
\epsscale{1}
\plotone{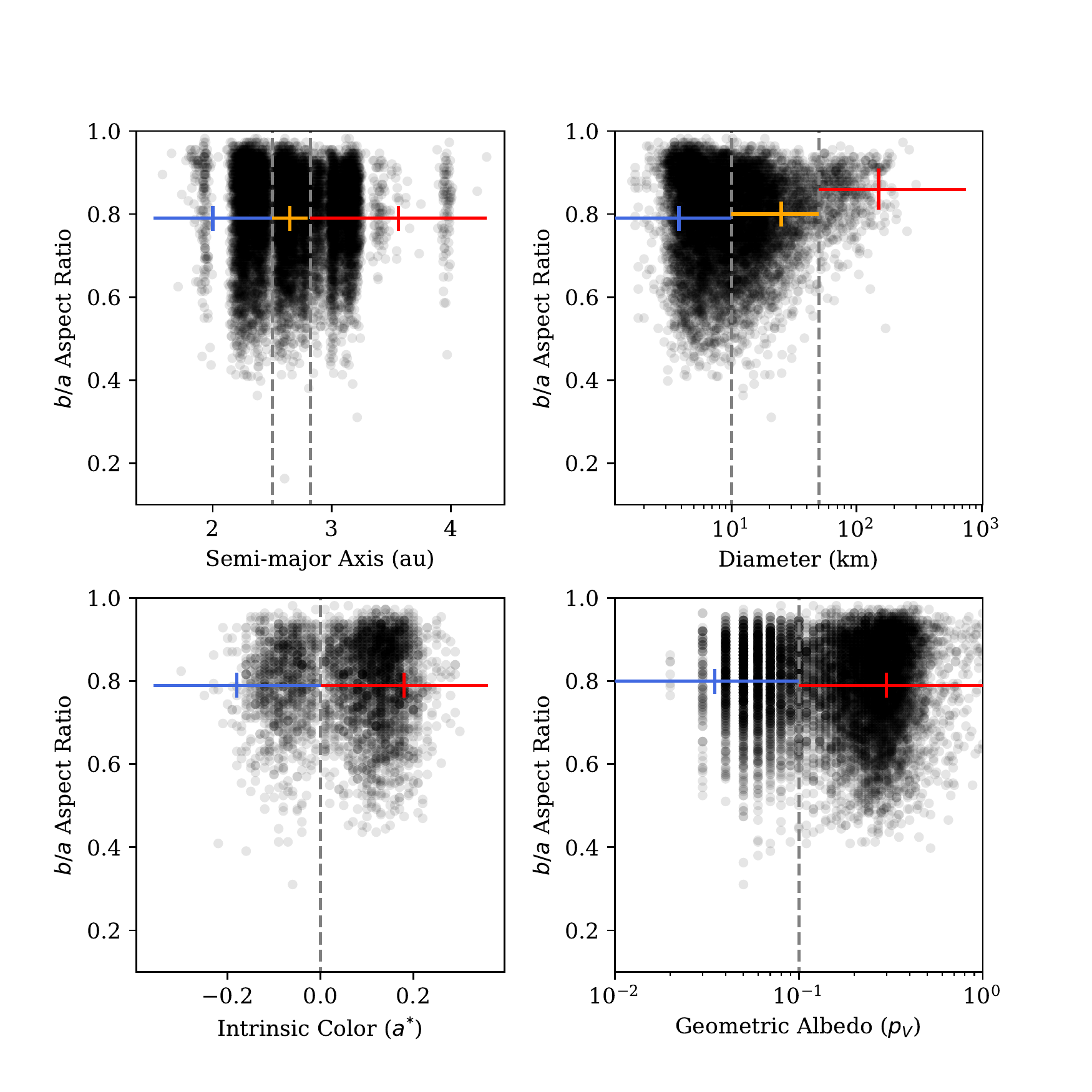}
\caption{Correlations of the $b/a$ aspect ratios measured in this work
  with other target properties from the literature: semi-major axis
  (top left), target diameter (top right), intrinsic color (bottom
  left), and geometric albedo (bottom right). Semi-transparent dots
  represent individual objects. The colored error symbols
  represent the average $b/a$ aspect ratio per sub-population as
  indicated by the vertical dashed lines (compare to Table \ref{tab:subsamples}). \label{fig:correlations}}
\end{figure}

Matching objects and measured properties are listed in Table
\ref{tab:results} and plotted in Figure \ref{fig:correlations}. Aspect
ratios ($b/a$) have been derived from the
lower limit amplitudes $A$ using the relation
$b/a = 10^{-0.4\cdot A}$. We furthermore repeat the analysis performed
in Section \ref{sec:results_shapes} for sub-samples that are defined
based on the respective parameter distributions. The sub-sample
definitions and average aspect ratio results are listed in Table
\ref{tab:subsamples}. In order to derive the latter, we use the exact same model as described in Section \ref{sec:results_shapes} but applied to the individual sub-samples. 

\begin{deluxetable*}{lccrc}
\tablecaption{Sub-Sample Definitions and Average Aspect Ratios\label{tab:subsamples}}
\tablecolumns{5}
\tablewidth{0pt}
\tablehead{
\colhead{Parameter} &
\colhead{Range} &
\colhead{Description} &
\colhead{$N_{\mathrm{obj}}$\tablenotemark{a}} &
\colhead{Average Aspect Ratio ($b/a$)}}
\startdata
Semi-major axis ($a$) & $a < 2.5$~au & inner main belt & 4,054 & $0.79\pm0.03$ \\
& $2.5 \leq a < 2.82$~au & middle main belt & 3,880 & $0.79\pm0.03$ \\
& $a \geq 2.82$~au & outer main belt & 3,731 & $0.79\pm0.03$ \\
\hline
Intrinsic Color ($a^{\text{*}}$) & $a^{\text{*}} < 0$ & blueish asteroids & 1,008 & $0.79\pm0.03$ \\
& $a^{\text{*}} \geq 0$ & reddish asteroids & 2,547 & $0.79\pm0.03$ \\
\hline
Diameter ($d$) & $d < 10$~km & small asteroids & 5,821 & $0.79\pm0.03$ \\
& $10 \leq d < 50$~km & medium-sized asteroids & 3,673 & $0.80\pm0.03$ \\
& $d \geq 50$~km & large asteroids & 468 & $0.86\pm0.05$ \\
\hline
Geometric Albedo ($p_V$) & $p_V \leq 0.1$ & primitive asteroids & 2,827 & $0.80\pm0.03$ \\
& $p_V > 0.1$ & non-primitive asteroids & 7,135 & $0.79\pm0.03$ \\
\hline
--- & --- & all main belt asteroids & 11,665 & $0.80\pm0.04$\\
\enddata
\tablenotetext{a}{Number of objects in this subsample.}
\end{deluxetable*}

We find no significant differences in the average aspect ratios
derived from our model (see Section \ref{sec:results_shapes}) in
sub-samples based on intrinsic color, semi-major axis, and geometric albedo. % In
% the case of geometric albedo, there is an insignificant difference
% between primitive asteroids and non-primitive asteroids --- the former
% seem to be slightly rounder.
However, we find a difference in the average aspect ratio
  between medium-sized asteroids and large asteroids (see Table
  \ref{tab:subsamples} for definitions) at the $1\sigma$ level: our
  model suggests that main belt asteroids with diameters
${\geq} 50$~km are rounder than smaller asteroids, which
  agrees with trends from previous studies \citep{Cibulkova2016,
    Nortunen2017, Cibulkova2018}.

  We also use a $k$-sample Anderson-Darling test ({\tt
    scipy.stats.anderson\_ksamp}) to investigate differences in the
  amplitude distributions in the individual sub-samples. We find the
  null-hypothesis that the different subsamples in semi-major, object
  diameter, intrinsic color, and object albedo come from the same
  distribution can be rejected at the 1\% level. This indicates clear
  differences in the amplitude distributions as a function of these
  parameters, which do not affect the average aspect ratios derived
  from our model (see Figure \ref{fig:correlations}).
% The lack of {\bf unanimous}
% correlation in our data is in part caused by the small number of
% observations available in DR2.
% There seems to
% be a low--amplitude trend with semi-major axis $a > 3.3$~au (Figure
% \ref{fig:correlations}, top left), but a detailed study of this region
% was not possible due to the small sample size. 
More detailed studies
will be possible with the availability of a better coverage of the asteroid population with Gaia Data Release 3.

% All these results are supported by a running average
% analysis over our sample (green lines in Figure
% \ref{fig:correlations}), providing very similar results. 

\section{Discussion}
\label{sec:discussion}

\subsection{Comparison to previous work}
\label{sec:previouswork}

We compare our derived mean aspect ratio for main belt
  asteroids ($b/a=0.80\pm0.04$) with previous results. 
% derivations by \citet{McNeill2016},
%   \citet{Cibulkova2016}, \citet{Nortunen2017}, and
%   \citet{Cibulkova2018}. 
  \citet{McNeill2016} derived the average triaxial shape parameters of
  main belt asteroids with diameters ${\leq}20$~km from Pan-STARRS1
  observations as ($1:0.85\pm 0.13:0.71\pm 0.12$; this value has been
  updated to $1:0.83\pm 0.07: 0.70\pm 0.06$ with the updated form of
  the model which now more accurately accounts for phase angle
  amplitude correction), which agrees within the uncertainties with
  our findings for the same size regime. \citet{Cibulkova2016} find
  aspect ratios $b/a={\sim}0.63$ for main belt asteroids with
  diameters smaller than 25~km and $b/a={\sim}0.77$ for main belt
  asteroids greater than 50~km from the Lowell photometric database
  \citep{Bowell2014}. Both results are lower than our estimates --
  implying a higher average elongation of the targets -- although the
  general trend of larger objects being rounder is consistent with our
  results. This effect might be caused by the large individual
  photometric uncertainties (0.1--0.2~mag) of the data sample used in
  that work, which might overestimate lightcurve amplitudes
  \citep{Cibulkova2018}. \citet{Nortunen2017} use dense photometric
  observations \citep[e.g., from WISE,][]{Wright2010}, finding
  probability density functions for the triaxial shape parameters that
  agree with our result for asteroids with diameters greater than
  50~km, but they also find asteroids smaller than 50~km to be more
  elongated ($b/a{\sim}0.55$) than in this work. This discrepancy
  might point to an albedo-based selection bias in our sample of
  optically observed asteroids that is not existent in the WISE sample
  that affects this measurement. Detailed modeling of this bias and
  the underlying elongation distributions will be necessary to confirm
  this hypothesis, which we defer to the future when we can make use
  of a larger number of Gaia-observed main belt asteroids.  Finally,
  based on a prolate spheroidal shape model ($a \geq b = c$) and
  Pan-STARRS1 photometric data, \citet{Cibulkova2018} find
  $b/a\sim0.8$ for main belt asteroids with diameters greater than
  12~km, which also agrees with our result.

% The obtained mean axis ratio of $b/a=0.80\pm 0.05$ is in agreement
% with {\bf some}  previous studies of main belt shape distributions
% \citep{McNeill2016, Cibulkova2018},{\bf and in disagreement with some other studies \citep{Cibulkova2016, Nortunen2017}. 

 % \citet{McNeill2016} report an
% average triaxial shape $1:0.85\pm 0.13:0.71\pm 0.12$; this value has
% been updated to $1:0.83\pm 0.07: 0.70\pm 0.06$ with the updated form
% of the model which now more accurately accounts for phase angle
% amplitude correction. \citet{Cibulkova2018} assume all objects to be
% prolate spheroids (i.e., $a \geq b = c$) and find an average axis ratio
% $b/a=0.8$, consistent with our result. Assuming the objects to be
% prolate spheroids \citet{McNeill2017} also show an average axis ratio
% of $b/a\approx 0.8$. 

\subsection{Validation of our approach}
\label{sec:validation}

\begin{figure}
\epsscale{0.7}
\plotone{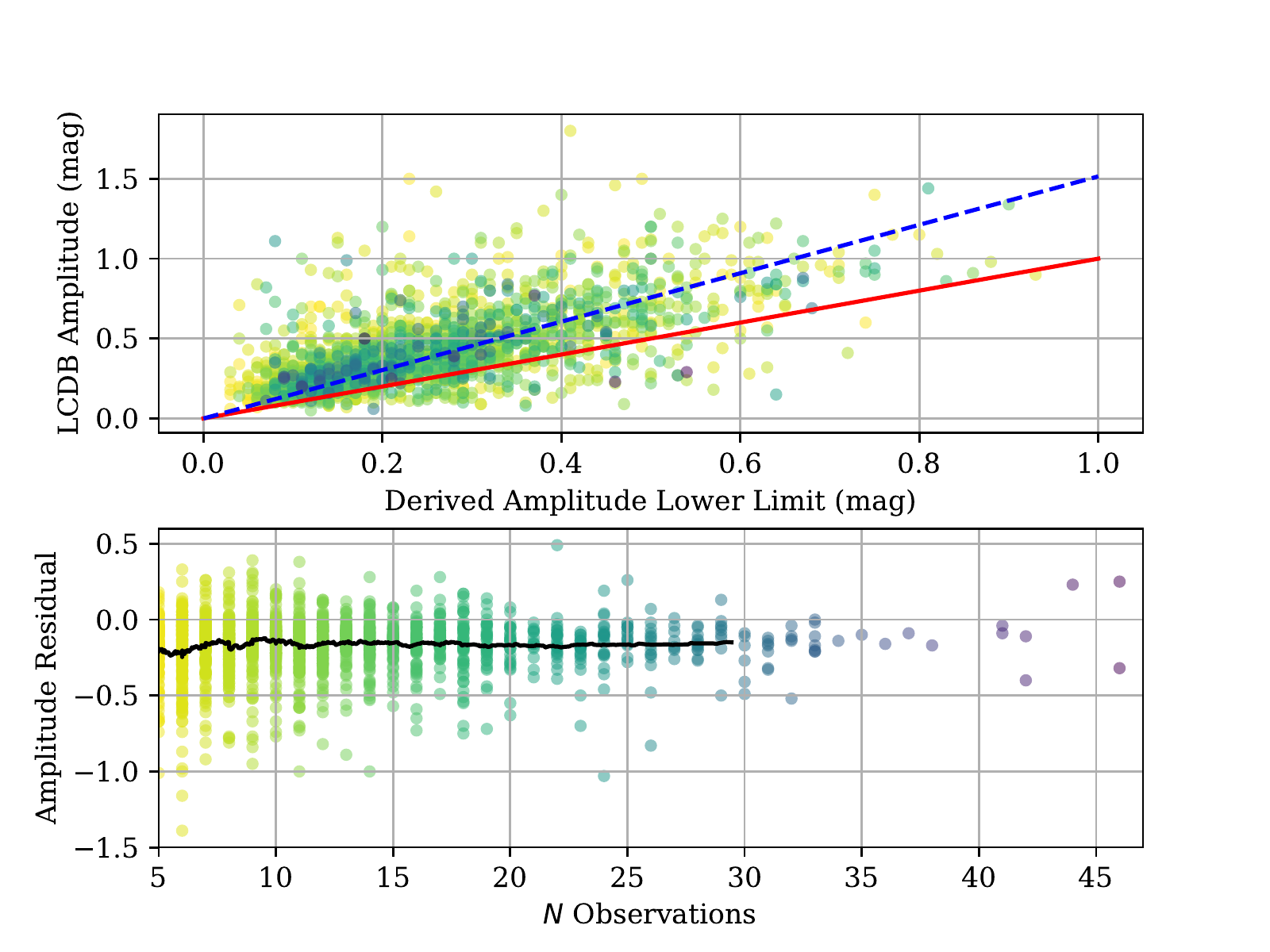}
\caption{{\bf Top}: Comparison of derived lower-limit lightcurve
  amplitudes and previously derived lightcurve amplitudes from the
  Asteroid Lightcurve Database \citep[LCDB,][]{Warner2009}. The red
  line indicates unity; the blue dashed line represents a linear fit
  to the scatter cloud. The colors of the points corresponds to the
  color scheme used in the bottom panel. {\bf Bottom}: Amplitude
  residuals (this work minus LCDB) as a function of the number of
  observations per target. The black line indicates a running average
  over 100 adjacent data points. \label{fig:lcdb}}
\end{figure}

We compare our lower-limit amplitudes against previously measured
amplitudes cataloged in the Asteroid Lightcurve Database
\citep[LCDB,][]{Warner2009}, restricting ourselves to the most reliable
information from that source (uncertainty parameter $U$=3).  The
amplitude comparison is plotted in the top panel of Figure
\ref{fig:lcdb}.  We find a significant fraction of small-amplitude
($A_{\mathrm{LCDB}}\leq$0.25~mag) targets to have overestimated amplitudes. This
effect is at least in part caused by incorrect phase slope parameters
$G_{12}$, which can lead to an artificial sloping of the derived $H_G$
magnitudes as a function of solar phase angle and hence lead to an
overestimation of the lightcurve amplitude.
% In order to quantify this effect, we investigate the potential
% slope differences over the range of measured phase slope parameters
% $-0.25 < G < 0.8$ \citep{Veres2015}. Over the average phase angle
% range (6\degr) centered on the average phase angle (19\degr), we find
% that this artificial slope can add up to 0.25~mag to the measured
% amplitude (indicated by the gray dashed line in Figure
% \ref{fig:lcdb}). 
Additional effects may include amplitude modulations as a function of
a target's observer-centric ecliptic longitude \citep[see][and Section
\ref{sec:limitations} for a discussion]{Bowell2014}. For larger
amplitudes ($A_{\mathrm{LCDB}}>0.25$~mag) the comparison shows a clear
underestimation of large amplitudes on our side. This effect is caused
by the incomplete sampling of the target's lightcurve. We quantify
this underestimation by fitting a linear slope to the data shown in
Figure \ref{fig:lcdb} (top). Based on this fit, we find that we
underestimate the lightcurve amplitudes by 51\%; the scatter of the
data points with respect to this slope is 20\%, providing a measure of
uncertainty for an individual object after applying the average 51\%
correction. We use these findings in Section \ref{sec:results_shapes}
to correct our derived lower-limit lightcurve amplitudes. The bottom
panel of Figure \ref{fig:lcdb} shows that this offset is independent
of the number of observations per target. However, we find amplitudes
derived from many observations to be more reliable.

\subsection{Limitations of this Analysis}
\label{sec:limitations}

This analysis is limited by a number of simplifications.
In our derivation of the lightcurve
amplitudes of our targets (Section \ref{sec:phasecurve}) we ignore the
effect of the solar phase angle on the lightcurve amplitude. From the
limited amount of DR2 observations, we are unable to account for this
effect on a per-object basis. Instead, we investigate the magnitude of
this effect on the ensemble of measured lightcurve amplitudes using
the relation derived by \cite{Zappala1990} and an assumed average
phase slope parameter of 0.02~mag/degree together with the average
solar phase angle in our sample (19\degr). Based on these assumptions,
we expect our lightcurve amplitudes to be overestimated by
40\%. However, in Section \ref{sec:validation}, our comparison of the
lightcurve amplitudes derived in this work with previously measured
lightcurve amplitudes from the literature shows that we are on average
underestimating lightcurve amplitudes. We interpret this discrepancy
as a sign that the effect of the solar phase angle on the lightcurve
amplitude is insignificant for our ensemble analysis.

Furthermore, the spin pole orientations of the majority of our sample
targets are unknown. For individual objects, this lack of information
can be critical in two ways: (1) the relative orientation of the spin
pole axis to the observer directly modulates the observed lightcurve
amplitude in a geometrical effect, and (2) the orientation of the spin
poles axis introduces additional variability that can exceed the
target's lightcurve amplitude as a function of the observer-centric
ecliptic longitude \citep[see, e.g.,][]{Bowell2014}.

We justify the use of the spin pole distribution derived for asteroids
smaller than 30~km \citep{Hanus2011} in our model (Section
\ref{sec:results_shapes}) with the fact that the majority of our
sample objects with measured diameters fall into this size regime
(${\sim}$90\%). Nevertheless, we investigate the impact of this choice
on the results and redo the analysis in Section
\ref{sec:results_shapes} based on a uniform distribution of spin poles
in both ecliptic latitude and longitude. This distribution agrees with
the distribution for larger asteroids \citep{Hanus2011} and represents
the extreme opposite to our original choice, allowing us to probe any
differences in the results. The resulting average aspect ratios for
all samples agree with our nominal results within their 1$\sigma$
uncertainties. This observation is supported by the results of a
Kolmogorov-Smirnov test \citep[e.g.,][]{Press1992} in which we compare
the de-biased amplitude distributions generated in our model based on
the different spin pole distributions and find no significant
differences between those distributions.

Finally, we assume in our derivation of the average aspect ratios that
the $b$ and $c$ axis ratios of our asteroids will be equal. Shape
models from dense light curve inversion are the only existing data
sets which can detail the relationship between these axes. These data,
however, are incomplete and likely to be biased toward higher $b/c$
values. Allowing $b/c$ to vary will slightly inflate the value
obtained for the $b/a$ axis ratio but not to an extent greater than
the uncertainties produced by the model.

\section{Conclusions}
\label{sec:conclusions}

We measure lower limit lightcurve amplitudes for 11,665
main belt asteroids in the Gaia Data Release 2 catalog. We 
  derive the mean aspect ratio for main belt asteroids as
$b/a=0.80\pm0.04$, which is in agreement with previous studies. We
investigate trends in the shape distribution as a function of
semi-major axis, intrinsic color, diameter, and geometric albedo,
using data from the literature. Based on our model simulations,
  we find that main belt asteroids greater than 50~km in diameter
have on average higher $b/a$ aspect ratios (are rounder) than smaller
asteroids. We furthermore find significant differences in the
  derived lower limit amplitude distributions with respect to
  semi-major axis, intrinsic color ($a^*$), and geometric albedo.  We
predict that more detailed population and shape distribution studies
will be possible with the availability of Gaia Data Release 3.

\acknowledgments

The authors would like to thank an anonymous referee for a
  thorough review that improved the quality of this work
  significantly.  This work is supported in part by NSF award
1229776.  This work has made use of data from the European Space
Agency (ESA) mission {\it Gaia}
(\url{https://www.cosmos.esa.int/gaia}), processed by the {\it Gaia}
Data Processing and Analysis Consortium (DPAC,
\url{https://www.cosmos.esa.int/web/gaia/dpac/consortium}). Funding
for the DPAC has been provided by national institutions, in particular
the institutions participating in the {\it Gaia} Multilateral
Agreement.  This research has made use of the VizieR catalogue access
tool, CDS, Strasbourg, France. The original description of the VizieR
service was published in A\&AS 143, 23


\vspace{5mm}
\facility{Gaia}


\software{astropy \citep{astropy}, astroquery
  (\url{https://github.com/astropy/astroquery}), matplotlib
  (\url{https://matplotlib.org/}), numpy
  (\url{http://www.numpy.org/}), scipy (\url{https://scipy.org}),
    sbpy (\url{https://sbpy.org}), topcat
  (\url{http://www.star.bris.ac.uk/~mbt/topcat/}) }



%\appendix

%\section{Appendix information}



\end{document}